\documentclass[12pt]{iopart}
\usepackage{iopams,graphicx}
\newfont{\bfrak}{eufb10 at 12pt}
\def\wb{{\overline W}}
\def\bZ{{\bf Z}}
\def\bX{{\bf X}}
\begin{document}

\title[Eigenvectors in the Superintegrable Model I]%
{Eigenvectors in the Superintegrable Model I:\break
${\frak{sl}}_2$ Generators}

\author{Helen Au-Yang and Jacques H H Perk}

\address{Department of Physics, Oklahoma State University, %\\
145 Physical Sciences, Stillwater, OK 74078-3072, USA}
\ead{perk@okstate.edu, helenperk@yahoo.com}
\begin{abstract}
In order to calculate correlation functions of the chiral Potts
model, one only needs to study the eigenvectors of the
superintegrable model. Here we start this study by looking for
eigenvectors of the transfer matrix of the periodic $\tau_2(t_q)$ model
which commutes with the chiral Potts transfer matrix.
We show that the degeneracy of the eigenspace of $\tau_2(t_q)$ in the
$Q=0$ sector is $2^r$, with $r=(N\!-\!1)L/N$ when the size of the
transfer matrix $L$ is a multiple of $N$. We introduce chiral Potts
model operators, different from the more commonly used generators of
quantum group $\tilde{\mathrm U}_q({\widehat{\frak{sl}}}_2)$. From
these we can form the generators of a loop algebra
$\mathcal{L}({\frak{sl}}_2)$. For this algebra, we then use
the roots of the Drinfeld polynomial to give new explicit
expressions for the generators representing the loop algebra
as the direct sum of $r$ copies of the simple algebra
${\frak{sl}}_2$.
\end{abstract}

\vspace{2pc}
%\noindent{\it Keywords}: 
%\maketitle

%%%%%%%%%%%%%%%%%%%%%%%%%%%%%%%%%%%%%%%%%%%%%%%%%%%%%%%%%%%%%%%%%%%%%%%%%

The integrable chiral Potts model is an $N$-state spin model on a
planar lattice, whose Boltzmann weights require high-genus algebraic
functions for their parameterization~\cite{AMPTY,BPTS,AMT,BPA,APTa}.
Nevertheless much progress has been made. The model has very special
properties which made it possible for Baxter to calculate the free
energy and order parameters~\cite{BaxS,BaxO1,BaxO2,BaxO}. It seems
likely that correlation functions of this model can also be
calculated.

Unlike the calculation of the free energy and order parameters, for
which the knowledge of the eigenvalues of the transfer matrix is
sufficient, to calculate the correlation functions, we also need
information about the eigenvectors. We shall show that it is only
necessary to study the eigenvector space in the superintegrable model,
which is in many ways similar to the Ising model~\cite{Onsager,Yang}.
Particularly, one can construct a loop algebra in the superintegrable
model similar to the Onsager algebra~\cite{Onsager} in the Ising
model.

%------------------------------------------------------------------------
\begin{figure}[tbh]
\begin{center}
\includegraphics[width=\hsize]{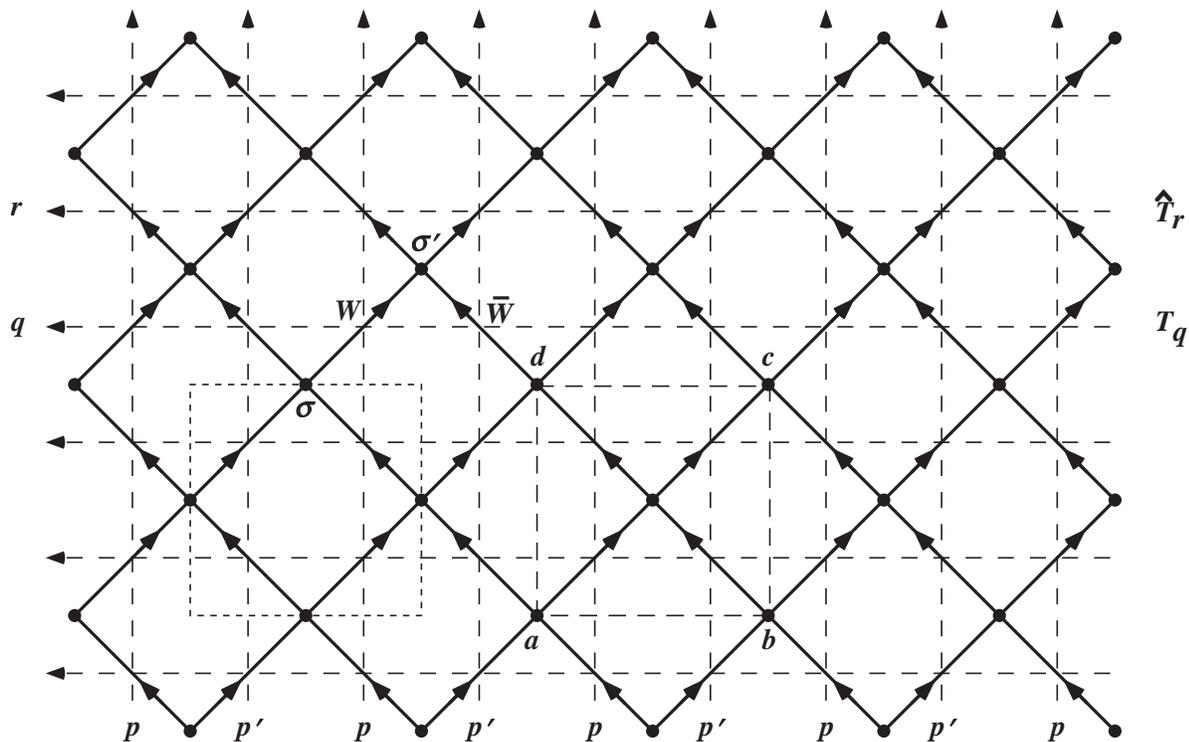}
\end{center}

\caption{The oriented lattice of a two-dimensional spin model
is represented by solid lines. At each lattice site, indicated by
a black dot, there is a spin which can take $N$ different values. The
Boltzmann weights $W$ and $\wb$ are associated with pair interactions
along the edges. The medial graph consists of oriented
dashed lines---the ``rapidity lines"---carrying variables
$p,q,\ldots$. To each rapidity line $q$ (or $r$), we associate a
transfer matrix $T_q$ (or ${\hat T}_r$) and these transfer matrices
commute with one another for a $Z$-invariant lattice. The product of
two transfer matrices $T_q{\hat T}_r$ can be written as products of
``stars" $\mathcal{U}(a,b,c,d)$, one of which is drawn inside the dashed
box; assuming periodic boundary conditions, it can also be written as
a trace of the product of intertwiners $\mathcal{S}$, ``squares," shown
to the left inside the dotted box.}
\end{figure}
%--------------------------------------------------------------------------
We consider here the square lattice drawn diagonally with edges
denoted by solid lines as in Fig.~1. Each spin $\sigma$ may take $N$
different values, and interacts with each neighboring spin $\sigma'$.
Boltzmann weights $W_{pq}(\sigma-\sigma')$ (or
$\bar{W}_{pq}(\sigma-\sigma')$) are associated with pair interactions
along the SW-NE edges (or the SE-NW edges). The rapidity variables
$p$ and $q$ are denoted by the dashed oriented lines in Fig.~1. The
Boltzmann weights are given in product forms as
\begin{equation}
W_{pq}(n)={\Bigl(\frac{\mu_p}{\mu_q}\Bigr)}^{\!n}
\prod^{n}_{j=1}\frac{y_q-x_p\omega^j}{ y_p-x_q\omega^j},\quad
\wb_{pq}(n)={(\mu_p \mu_q)}^n
\prod^{n}_{j=1}\frac{\omega x_p-x_q\omega^j}{ y_q-y_p\omega^j}.
\end{equation}
where $\omega^N=1$, $W_{pq}(N+n)=W_{pq}(n)$ and
$\wb_{pq}(N+n)=\wb_{pq}(n)$. To each horizontal and vertical rapidity
line
$p$, two variables $x_p$ and $y_p$ are assigned and they are related
by the equations\footnote{Korepanov introduced a version of the
$\tau_2$-model in 1986 and discovered the genus $(N-1)^2$ curve
in the second member of (\ref{curve}) as its ``vacuum curve"
using Baxter's ``pair-propagation through a vertex" method to find
the condition under which the transfer matrices commute.
Korepanov's work was not known to us when we discovered the genus-10
curve for the 3-state chiral Potts model late in 1986~\cite{AMPTY},
as it became available in the West only seven years
later~\cite{Korepanov}.}
\begin{equation}
\mu_p^N\!=\!k'/(1-k\,x_p^N)\!=\!(1-k\,y_p^N)/k',\quad
x_p^N+y_p^N\!=\!k(1+x_p^N y_p^N),
\label{curve}
\end{equation}
${k\vphantom{'}}^2+{k'}^2\!=\!1$, which determine a high-genus curve.
The curve is symmetric in $x_p$ and $y_p$, and there exist various
automorphisms, such as $p\to Rp$, $p\to Up$,
\begin{equation}
(x_{Rp},y_{Rp},\mu_{Rp})=(y_p,\omega x_p,1/\mu_p),\quad
(x_{Up},y_{Up},\mu_{Up})=(\omega x_p,y_p,\mu_p),
\end{equation}
which leave the curve invariant.

To each horizontal rapidity line $q$, $r$ one associates a transfer
matrix, which is the product of all the weights,
\begin{eqnarray}
T_q=T(x_q,y_q)_{\sigma \sigma'}=\prod_{J=1}^{L}
W_{p q}(\sigma^{\vphantom{'}}_J -\sigma'_J)
\wb_{p'q}(\sigma^{\vphantom{'}}_{J+1} -\sigma'_{J}),
\nonumber\\
{\hat T}_r={\hat T}(x_r,y_r)_{\sigma' \sigma''}=
\prod_{J=1}^{L}\wb_{pr}(\sigma'_J -\sigma''_J)
W_{p'r}(\sigma'_J -\sigma''_{J+1}).
\end{eqnarray}
When the rapidities of the two transfer matrices $T_q$ and $\hat T_r$
are related by $x_r=y_q$ and $y_r=\omega^j x_q$, the product of these
two transfer matrices decomposes~\cite{BBP}, i.e.
\begin{equation}
T_q{\hat T}_r\to\bar h_j\tau_j(t_q)+h_j\tau_{N-j}(\omega^j
t_q),\quad t_q=x_q y_q.
\end{equation}
Furthermore, as the matrices $\tau_j(t_q)$ still
commute with the transfer matrices, they satisfy additional
equations. These functional relations~\cite{BBP} were used by
Baxter~\cite{BaxS,BaxO} to calculate the free energy and the order
parameters.

It is known that $\tau_2(t_q)$ can be written as a trace of the product
of $\cal L$ matrices which is defined by what is in the dotted box in
Fig.~1,
\begin{equation}
\tau_2(t_q)=
\Tr[\mathcal{L}_L(t_q)\cdots\mathcal{L}_2(t_q)\mathcal{L}_1(t_q)].
\label{tauL}\end{equation}
Bazhanov and Stroganov~\cite{BS} have shown that these $\cal L$
matrices satisfy the Yang--Baxter equation
$\mathcal{R}\mathcal{L}\mathcal{L'}=\mathcal{L'}\mathcal{L}\mathcal{R}$,
in which $\mathcal{R}$ is the six-vertex $R$-matrix in the vertex language.
This shows that the eigenspace of $\tau_2(t_q)$ is related to the
representations of the quantum group
$U_q({\widehat{\hbox{$\frak{sl}$}}}_2)$.

Since the chiral Potts model satisfies the star-triangle equation,
rapidity lines can be moved through vertices without changing
the partition function $Z$, whence such models are called
$Z$-invariant~\cite{Bax-ZI}. Consequently, the order parameters
depend only on the temperature variable $k$, whereas a pair
correlation function depends only on $k$ and the rapidity lines
sandwiched between the two spins. Because of this, the
diagonal correlation function of two spins separated in the vertical
direction in Fig.~1 depends only on the horizontal rapidities lines and
is independent of $p$ and $p'$, which we may choose such that
$x_p'=y_p$ and $y_p'=x_p$ implying that the model becomes
superintegrable. On the other hand, the diagonal correlations
in the horizontal direction are functions of
$p$ and $p'$ only. Thus they are intimately related to correlation
functions of $\tau_j(t_q)$. For $j=2$, especially for certain open
boundary conditions, some work has recently been done on the
eigenvectors of $\tau_2(t_q)$, see e.g.~\cite{Iorgov,GIPS,GIPST}.
Here we shall use a different approach in order to get some more
understanding of the consequences of superintegrability.

For the superintegrable case, $\tau_2(t_q)$ has simple eigenvalues
given by~\cite{Baxsu,BaxIf1,Tara}
\begin{equation}
\tau_2(t_q) \nu_Q=
[(1-\omega t)^L+\omega^{-Q}(1-t)^L]\nu_Q,\quad Q\in {\mathbb Z}_N,\quad
t=t_q/t_p,
\label{eigen2}\end{equation}
where the $\nu_Q$ are common eigenvectors for all $\tau_j(t_q)$,
the product of two transfer matrices and also the spin shift operator
$\mathcal{X}$, which shifts all spins by 1, i.e.\ $\sigma_j\to\sigma_j+1$.
The functional relations are matrix equations, but as these matrices
commute, Baxter treated them as relations between the eigenvalues;
that is he treated these relations as scalar equations. For the
correlations, one needs information about the eigenvectors, and we must
treat the functional relations as matrix identities as Tarasov
did~\cite{Tara}, that is, the matrices are expressed in terms of the
eigenvectors of the spin shift operator $\mathcal{X}$.

The eigenvectors of $\mathcal{X}$ are
\begin{equation}
|Q;n_1,\cdots,n_L\rangle=|Q;\{n_j\}\rangle=
N^{-{\frac12}}\sum_{\sigma_1=0}^{N-1}\omega^{-Q\sigma_1}
|\sigma_1,\sigma_2,\cdots,\sigma_L\rangle,
\label{maps}\end{equation}
where $n_j=\sigma_j-\sigma_{j+1}$ is the difference between the adjacent
spins and can be considered as variables associated to the $j$th edge.
The cyclic boundary condition $\sigma_{L+1}=\sigma_{1}$ becomes
$n_1+\cdots+n_L=0\,\hbox{(mod $N$)}$. Obviously,
$\mathcal{X}|Q;\{n_j\}\rangle=\omega^{Q}|Q;\{n_j\}\rangle$.

In the superintegrable case, the $N$ functional relations
\begin{equation}
T_Q(x_q,y_q){\hat T}_Q(y_q,\omega^j x_q)\nu_Q=
C\,\mathcal{P}(t)\nu_Q\,,\quad t=t_q/t_p,
\label{tbt}\end{equation}
are all the same, independent of $j$, where $C$ is a factor collecting
the poles of the left-hand side~\cite{BaxIf1}. The transfer matrix,
which depends only on the difference $\ell=\sigma_1-\sigma'_1$ in the
space of the eigenvectors of the spin shift operator $\mathcal{X}$, is
\begin{equation}
\langle\{n_j'\}|T_Q(x_q,y_q)|\{n_j\}\rangle=\langle
Q;\{n'_j\}|T_q|Q;\{n_j\}\rangle=
\sum_{\ell=0}^{N-1}\omega^{-Q\ell}
(T_q)_{\sigma,\sigma'}
\label{Tq}\end{equation}
which is the Fourier transform over $\ell$, whereas
\begin{equation}
\mathcal{P}(t)= P(t^N)=
\frac{t^{-Q}}N
\sum_{n=0}^{N-1}\omega^{-nQ}\frac{(1-t^N)^L}{(1-\omega^n t)^L}.
\label{Poly}\end{equation}
We shall consider the case $Q=0$ only. For other cases,
see~\cite{BaxIf1} for details. The function $\mathcal{P}(t)$ is a
polynomial in $z=t^N$ and is of order $r=(N-1)L/N$. From (\ref{tbt})
and the $r$ zeros of (\ref{Poly}), Baxter obtained $2^r$ eigenvalues of
the transfer matrix $T_q$, and thus found the free energy. What is
implicit in Baxter's calculation is that corresponding to the $Q=0$
eigenvalue in (\ref{eigen2}), there are $2^r$ eigenvectors; the
eigenspace of $\tau_2(t_q)$ is highly degenerate. For $N=3$ and small
number of sites $L$, we can calculate all the eigenvalues and
eigenvectors of $\tau_2(t_q)$ and find that the degeneracy is $2^r$ with
$r=(N-1)L/N$ only if $L$ is a multiple of $N$. These results agree with
those of Deguchi~\cite{Degu}.

Instead of (\ref{tauL}), we let $\tau_2(t_q)$ be written as a product of
$\mathcal{U}(a,b,c,d)$, which is what is inside the dashed box in
Fig.~1,
\begin{equation}
\tau_2(t_q)=
\prod_{J=1}^L\mathcal{U}(\sigma_J,\sigma_{J+1},\sigma'_{J+1},\sigma'_J).
\label{tauU}\end{equation}
It is easily shown that the Yang--Baxter equation
$\mathcal{RUU'}=\mathcal{U'UR}$ also holds, but now it is in the IRF
language.

For the superintegrable case, the only nonvanishing elements of
$\cal U$ are
\begin{eqnarray}
\mathcal{U}(a,b,b,a)=1-\omega^{n+1}t,\quad \mathcal{U}(a,b,b-1,a)=
\omega t(\omega^{n+1}-1),\nonumber\\ 
\mathcal{U}(a,b,b,a-1)=(1-\omega^{n}),\quad\mathcal{U}(a,b,b-1,a-1)=
\omega (\omega^{n}-t),
\label{U}\end{eqnarray}
where $n=a-b$, which is one of the edge variables.
We may write
\begin{equation}
\mathcal{U}(a,b,c,d)=\mbox{\bfrak u}(a-d, b-c)_{d-c,a-b}.
\label{Uu}\end{equation}
Using the usual convention
\begin{eqnarray}
\bZ_{n,m}=\langle n|\bZ|m\rangle=\omega^m\delta_{n,m},\quad
&\bZ|m\rangle=\omega^m|m\rangle,\nonumber\\
\bX_{n,m}=\langle n|\bX|m\rangle=
\delta_{n,m+1},\quad&\bX|m\rangle=|m+1\rangle,
\end{eqnarray}
we find from (\ref{U}) and (\ref{Uu}) that
\begin{eqnarray}
\mbox{\bfrak u}(0, 0)=(1-\omega t\bZ),\quad
\mbox{\bfrak u}(0, 1)=-\omega t(1-\bZ)\bX,\nonumber\\
\mbox{\bfrak u}(1, 0)=\bX^{-1}(1-\bZ),\quad
\mbox{\bfrak u}(1, 1)=\omega(\bZ-t).
\label{u}\end{eqnarray}
Due to the way the arrows on the rapidity lines are drawn in Fig. 1,
corresponding to the choice in the original paper~\cite{BPA}, the
matrix multiplication is either down to up and right to left, or
from left to right and up to down. We use the latter choice, such
that the ket vectors $|\;\rangle$ are related to the variables
$\{n_j\}$ on the lower edges of the $L$ faces. Letting
\begin{eqnarray}
\bX_j={\bf 1}\otimes\cdots\otimes{\bf 1}\otimes
\hbox{\vbox{\hbox{\;$_j$}\hbox{$\bX$}}}\otimes
{\bf 1}\otimes\cdots\otimes{\bf 1},\nonumber\\
\bZ_j={\bf 1}\otimes\cdots\otimes{\bf 1}
\otimes\hbox{\vbox{\hbox{\;$_j$}\hbox{$\bZ$}}}\otimes
{\bf 1}\otimes\cdots\otimes{\bf 1},
\end{eqnarray}
we may define
\begin{equation}
(1-\omega)\mbox{\bfrak e}_j=\bX_j^{-1}(1-\bZ_j),\quad
(1-\omega)\mbox{\bfrak f}_j=(1-\bZ_j)\bX_j,
\label{ef}\end{equation}
such that
\begin{equation}
(1-\omega)(\mbox{\bfrak e}_j\mbox{\bfrak f}_j-
\omega\mbox{\bfrak f}_j\mbox{\bfrak e}_j)
=(1-\omega\bZ_j^2),
\label{cef}\end{equation}
which relations are not the same as the ones for the cyclic and
nilpotent representation of $U_q({\frak{sl}}_2)$ used by
Jimbo~\cite{JimboNK,NiDe}. Nevertheless, it is easy to show that
$\mbox{\bfrak e}^N=0$ and $\mbox{\bfrak f}^N=0$, and
\begin{eqnarray}
\mbox{\bfrak e}|0\rangle=0,\quad \mbox{\bfrak e}|n\rangle=
[n]|n-1\rangle,\nonumber\\
\mbox{\bfrak f}|N-1\rangle=0,\quad
\mbox{\bfrak f}|n\rangle=[n+1]|n+1\rangle,
\label{efon}\end{eqnarray}
where $[n]=(1-\omega^n)/(1-\omega)=1+\omega+\cdots+\omega^{n-1}$.
This definition, though different from the one more commonly used in
quantum groups, is not new in the literature. This $[n]$ is also
a symbol defined in $q$-series~\cite{AAR}.

{}From (\ref{u}) and (\ref{ef}) we may associate to each face an
operator,
\begin{equation}
\mbox{\bfrak u}_j=\left[\begin{array}{cc}
(1-\omega t\bZ_j) &-\omega t(1-\omega)\mbox{\bfrak f}_j\cr
(1-\omega)\mbox{\bfrak e}_j&\omega(\bZ_j-t)\end{array}\right].
\label{uj}\end{equation}
By multiplying all these operators for the $L$ faces together, we
obtain
\begin{equation}
{\bf U}(t)=
\mbox{\bfrak u}_1\cdots \mbox{\bfrak u}_L=\left[\begin{array}{cc}
{\bf A}(t) &{\bf B}(t)\cr
{\bf C}(t) &{\bf D}(t)\end{array}\right],
\end{equation}
such that,
\begin{equation}
\tau_2(t_q)\big|_Q\equiv\langle Q;\{n'_j\}|\tau_2(t_q)|Q;\{n_j\}\rangle=
{\bf A}(t)+\omega^{-Q} {\bf D}(t),
\label{tauq}\end{equation}
for $Q=0,\cdots,N-1$. From (\ref{uj}), we can see easily that the elements
of ${\bf U}(t)$ are polynomials in $t$,
\begin{equation}
{\bf U}(t)=\sum_{j=0}^L\,(-\omega t)^j
\left[\begin{array}{cc}
{\bf A}_j &{\bf B}_j\cr
{\bf C}_j &{\bf D}_j\end{array}\right].
\end{equation}
A few of the coefficients of these polynomials are easy to find.
Particularly:
\begin{eqnarray}
{\bf A}_0={\bf D}_L={\bf 1},\quad
{\bf A}_L={\bf D}_0\,\omega^{-L}=\prod_{j=1}^L\bZ_j,
\quad {\bf C}_L={\bf B}_0=0;\label{AD}\\
{\bf B}_L=
(1-\omega)\sum_{j=1}^L\prod_{m=1}^{j-1}\bZ_m{\mbox{\bfrak f}}_j,\quad
{\bf C}_0=(1-\omega)\sum_{j=1}^L\omega^{j-1}\prod_{m=1}^{j-1}\bZ_m
{\mbox{\bfrak e}}_j,\nonumber\\
{\bf B}_1\!=
\!(1\!-\!\omega)\sum_{j=1}^L\,\omega^{L-j}{\mbox{\bfrak f}}_j
\!\!\prod_{m=j+1}^{L}\bZ_m,
\quad
{\bf C}_{L-1}\!=\!(1\!-\!\omega)\sum_{j=1}^L\,{\mbox{\bfrak e}}_j
\!\!\prod_{m=j+1}^{L}\bZ_m.
\label{BC}\end{eqnarray}
Each term in the operators $B_1$ and $B_L$ raises only one of the
$n_j$'s to $n_j+1$ and the resulting state does not satisfy the cyclic
boundary condition $\sum n_j=\ell N$; thus we need to consider the
product of $N$ of them. Define
\begin{eqnarray}
&&{\bf B}_j^{(N)}=\frac{{\bf B}_j^N}{[N]!},\quad
\hbox{for }j\!=\!1\hbox{ or }L,\nonumber\\
&&{\bf C}_n^{(N)}=\frac{{\bf C}_n^N}{[N]!},\quad
\hbox{for }n\!=\!0\hbox{ or }L\!-\!1,
\label{BCN}\end{eqnarray}
with $[N]!=[N]\cdots[2][1]$. More generally, we define
${\bf O}^{(n)}={\bf O}^n/{[n]!}$ for operator ${\bf O}$ and
$n=1,2,\cdots,N$.
{}From the Yang--Baxter equation, we find for $Q=0$,
\begin{eqnarray}
&[\tau_2(t_q),{\bf B}_L^{(N)}]=
(\omega-1){\bf B}(t){\bf B}_L^{(N-1)}({\bf A}_L-1),\nonumber\\
&[\tau_2(t_q),{\bf B}_1^{(N)}]=
(1-\omega^{-1})t^{-1}{\bf B}(t){\bf B}_1^{(N-1)}({\bf D}_0-1),
\nonumber\\
&[\tau_2(t_q),{\bf C}_0^{(N)}]=
(\omega-1){\bf C}(t){\bf C}_0^{(N-1)}({\bf D}_0-1),\nonumber\\
&[\tau_2(t_q),{\bf C}_{L-1}^{(N)}]=
(\omega-1)\omega t{\bf C}(t){\bf C}_{L-1}^{(N-1)}({\bf A}_L-1).
\label{comm}\end{eqnarray}
Let $\mathcal{S}$ be the set of all vectors
$|\psi\rangle=|n_1,\cdots,n_L\rangle$, which satisfy the cyclic
boundary condition $\sum n_j=\ell N$, then we find from (\ref{AD})
that ${\bf A}_L|\psi\rangle=|\psi\rangle$; if $L$ is a multiple of $N$
then ${\bf D}_0|\psi\rangle=|\psi\rangle$. Thus in $\mathcal{S}$ the
four operators (\ref{BCN}) commute with $\tau_2(t_q)$. Denoting
$|\Omega\rangle=|0,\cdots,0\rangle$ and
$|{\bar\Omega}\rangle=|N\!-\!1,\cdots,N\!-\!1\rangle$, which are the
``ferromagnetic" and ``antiferromagnetic" ground states, we find from
(\ref{U}) that
\begin{eqnarray}
&{\bf A}(t)|\Omega\rangle=(1-\omega t)^L|\Omega\rangle,\quad
{\bf D}(t)|\Omega\rangle=(1-t)^L|\Omega\rangle,\nonumber\\
&{\bf A}(t)|{\bar\Omega}\rangle=(1- t)^L|{\bar\Omega}\rangle,\quad
{\bf D}(t)|{\bar\Omega}\rangle=(1-\omega t)^L|{\bar\Omega}\rangle.
\label{eigenAD}\end{eqnarray}
Comparing with (\ref{eigen2}), we find that they are eigenvectors of
$\tau_2(t_q)$ in this degenerate eigenspace. The commutation relations
in (\ref{comm}) show that other eigenvectors can also be obtained by
operating the raising operators on $|\Omega\rangle$, or the lowering
operators on $|{\bar\Omega}\rangle$. It is also obvious that the
eigenspace is more degenerate when $L$ is a multiple of $N$.

We now let $L=\ell N$ and show the connection with the loop algebra
$\mathcal{L}({\frak{sl}}_2)$ using the notation of Drinfeld
\begin{eqnarray}
(1-\omega)^N{\bf x}_{0}^-={\bf B}_{L}^{(N)},\quad
(1-\omega)^N{\bf x}_{1}^-={\bf B}_{1}^{(N)},\nonumber\\
(1-\omega)^N{\bf x}_0^+={\bf C}_0^{(N)},\quad
(1-\omega)^N{\bf x}_{-1}^+={\bf C}_{L-1}^{(N)}.
\label{x}\end{eqnarray}
The generators of the loop algebra $\mathcal{L}({\frak{sl}}_2)$
are required to satisfy the following relations,
\begin{eqnarray}
&{\bf h}_0=[{\bf x}_0^+,{\bf x}_{0}^-]=
[{\bf x}_{-1}^+,{\bf x}_{1}^-], \label{h0}\\
&[{\bf h}_0,{\bf x}_{i}^-]=2{\bf x}_{i}^-,\quad
[{\bf h}_0,{\bf x}_{-i}^+]=-2{\bf x}_{-i}^+
\nonumber\\
&[{\bf x}_{-i}^+,[{\bf x}_{-i}^+,[{\bf x}_{-i}^+,{\bf x}_{j}^-]]]=0,
\quad
[{\bf x}_{i}^-,[{\bf x}_{i}^-,[{\bf x}_{i}^-,{\bf x}_{-j}^+]]]=0,\quad
i\ne j,
\label{serre}\end{eqnarray}
with $i,j=0,1$. From the relations (\ref{cef}) for the raising and
lowering operator in (\ref{ef}), which differ from those of the quantum
group, we can show that the operators in (\ref{BC}) do not satisfy the
Serre relation. Therefore, the proof used by the authors in~\cite{DFM}
to prove (\ref{serre}) for (\ref{x}) cannot be repeated here. To prove
these relations, we need $q$-series identities at root-of-unity, which
are not available in the literature.

However, the Yang--Baxter equation can be used to show that (\ref{h0})
holds in the sector $\mathcal{S}$ in which all states satisfy the
periodic boundary condition. We can also prove that the other
identities hold for certain states,
\begin{eqnarray}
&[{\bf x}_0^{+},[{\bf x}_0^{+},[{\bf x}_0^{+},{\bf x}_{1}^{-}]]]
({\bf x}_{1}^{-})^{\!(n)}\!|\Omega\rangle\!=\!0,\nonumber\\
&\quad\{[[{\bf x}_{-1}^{+},{\bf x}_{1}^{-}],
{\bf x}_{1}^{-}]\!-\!2{\bf x}_{1}^{-}\}
({\bf x}_{1}^{-})^{\!(n)}\!|\Omega\rangle\!=\!0,\nonumber\\
&[{\bf x}_{\!-1}^{+},[{\bf x}_{\!-1}^{+},
[{\bf x}_{\!-1}^{+},{\bf x}_0^{-}]]]
({\bf x}_{0}^{-})^{\!(n)}\!|\Omega\rangle\!\!=\!0,\nonumber\\
&\quad\{[[{\bf x}_{0}^{+},{\bf x}_{0}^{-}],
{\bf x}_{0}^{-}]\!-\!2{\bf x}_{0}^{-}\}
({\bf x}_{0}^{-})^{\!(n)}\!|\Omega\rangle\!\!=\!0.
\end{eqnarray}
We have used Maple to check if the identities in (\ref{serre}) hold
in $\mathcal{S}$ for small systems with $N=3$, $L=6$ and $N=4$, $L=8$.
For the former case, the set $\mathcal{S}$ consists of $3^5=243$ states,
and for all of them we have found that these identities hold. For the
latter case, for which there are $4^7=16384$ states in $\mathcal{S}$, we
have used a random number generator to pick up states randomly and to
verify that the identities indeed hold. From the large number of
checks that we have made, we conclude confidently that the conditions
in (\ref{serre}) hold for the set $\mathcal{S}$.

As a consequence, the loop algebra
\begin{equation}
{\bf h}_m=[{\bf x}_{m-\ell}^+,{\bf x}_{\ell}^-],\quad
{\bf x}_{m+\ell}^{\pm}=
\mp {\textstyle \frac12}[{\bf h}_m,{\bf x}_{\ell}^{\pm}],\quad
\ell,m\in\mathbb{Z},
\label{loop}\end{equation}
can be defined on the sector $\mathcal{S}$.
Furthermore, from (\ref{BC}), (\ref{BCN}) and (\ref{x}), and using
notations introduced by Deguchi~\cite{Degu}, we may calculate
explicitly
\begin{eqnarray}
({\bf x}_0^-)^{(n)}=
\sum_{ {\{0\le\nu_m\le N-1\}}\atop{\nu_1+\cdots+\nu_L=nN}}
\prod_{m=1}^{L}\frac{{\mbox{\bfrak f}}_j^{\nu_m}}{[\nu_m]!}
\bZ_m^{\sum_{\ell>m}\nu_{\ell}},\nonumber\\
({\bf x}_0^+)^{(n)}=
\sum_{ {\{0\le\nu_m\le N-1\}}\atop{\nu_1+\cdots+\nu_L=nN}}
\prod_{m=1}^{L}
\bZ_m^{\sum_{\ell>m}\nu_{\ell}}\, \frac{\omega^{m\nu_m}
{\mbox{\bfrak e}}_j^{\nu_m}}{[\nu_m]!},\nonumber\\
({\bf x}_1^-)^{(n)}=
\sum_{ {\{0\le\nu_m\le N-1\}}\atop {\nu_1+\cdots+\nu_L=nN}}
\prod_{m=1}^{L}\frac{\omega^{-m\nu_m}
{\mbox{\bfrak f}}_j^{\nu_m}}{[\nu_m]!}
\bZ_m^{\sum_{\ell<m}\nu_{\ell}},\nonumber\\
({\bf x}_{-1}^+)^{(n)}=
\sum_{ {\{0\le\nu_m\le N-1\}}\atop{\nu_1+\cdots+\nu_L=nN}}
\prod_{m=1}^{L}
\bZ_m^{\sum_{\ell<m}\nu_{\ell}}
\frac{{\mbox{\bfrak e}}_j^{\nu_m}}{[\nu_m]!},
\end{eqnarray}
where the summations are over the $L$ variables $\nu_m$ for
$m=1,\cdots,L$. These equations and (\ref{efon}) are used to find
\begin{eqnarray}
{\bf h}_{0}|\Omega\rangle=
({\bf x}_{-1}^+)({\bf x}_1^-)|\Omega\rangle=({\bf x}_{0}^+)
({\bf x}_0^-)|\Omega\rangle=-r|\Omega\rangle,\label{r}\\
({\bf x}_{0}^+)^{(n)}({\bf x}_1^-)^{(n)}|\Omega\rangle
=({\bf x}_{-1}^+)^{(n)}({\bf x}_0^-)^{(n)}|\Omega\rangle=
\Lambda_n|\Omega\rangle,
\label{Lambda}\\
{\bf h}_{0}|{\bar\Omega}\rangle=
-({\bf x}_1^-)({\bf x}_{-1}^+)|{\bar\Omega}\rangle=
-({\bf x}_0^-)({\bf x}_{0}^+) |{\bar\Omega}\rangle=
r|{\bar\Omega}\rangle,\label{br}\\
({\bf x}_1^-)^{(n)}({\bf x}_{0}^+)^{(n)}|{\bar\Omega}\rangle=
({\bf x}_0^-)^{(n)}({\bf x}_{-1}^+)^{(n)}|{\bar\Omega}\rangle=
\Lambda_n|{\bar\Omega}\rangle,\label{bLambda}
\end{eqnarray}
where $r=(N\!-\!1)L/N$. Equation (\ref{serre}) and the above results
differ from those in~\cite{NiDe}. For this reason, we give some
details of our calculation as
\begin{equation}
\Lambda_n=\sum_{ {\{0\le\nu_m\le N-1\}}\atop
{\nu_1+\cdots+\nu_L=nN}}\!\!1, \quad
\mathcal{Q}(t)=\prod_{m=1}^L\left(\sum_{\nu_m=0}^{N-1}t^{\nu_m}\right)=
\frac{(1-t^N)^L}{(1-t)^L},
\label{Q}\end{equation}
where we have inserted $t^{\nu_m}$ in each of the $L$ sums in
$\Lambda_n$ to arrive at $\mathcal{Q}(t)$. The condition
${\nu_1\!+\!\cdots\!+\!\nu_L\!=\!nN}$ means $\Lambda_n$ is the
coefficient of $t^{nN}$ in the expansion of $\mathcal{Q}(t)$. This way
we find
\begin{equation}
\Lambda_n=\Lambda_{r-n}=\sum_{m=0}^n(-1)^{m}\left({{L}\atop{m}}\right)
\frac{(L)_{nN-mN}}{(nN-mN)!}.
\end{equation}
Comparing (\ref{Q}) with (\ref{Poly}), we find that the polynomial in
the above equation is identical to the one used by Tarasov and Baxter.
According to the evaluation representation on the loop
algebra~\cite{Degu,ChPr}, the dimension of the eigenspace generated by
these operators is $2^r$. This can be seen as follows: In a similar
fashion as in~\cite{Degu}, we can show by induction
\begin{eqnarray}
({\bf x}_{0}^+)^{(n-1)}({\bf x}_1^-)^{(n)}|\Omega\rangle=
\sum_{j=1}^n{\bf x}_j^- \Lambda_{n-j}|\Omega\rangle,\nonumber\\
({\bf x}_{1}^-)^{(n-1)}({\bf x}_0^+)^{(n)}|{\bar \Omega}\rangle=
\sum_{j=1}^{n}{\bf x}_{j-1}^+ \Lambda_{n-j}|{\bar \Omega}\rangle.
\label{xpm}\end{eqnarray}
For $n>r$, its left-hand side vanishes; there are thus only $r$
independent ${\bf x}_j^-$ or ${\bf x}_j^+$. Particularly, for $n=r+1$,
we have
\begin{eqnarray}
\sum_{j=1}^{r+1}{\bf x}_j^- \Lambda_{r+1-j}|\Omega\rangle
=\sum_{j=0}^{r}{\bf x}_{j+1}^-\Lambda_{j}|\Omega\rangle=0,
\nonumber\\
\sum_{j=1}^{r+1} {\bf x}_{j-1}^+ \Lambda_{r+1-j}|{\bar \Omega}\rangle
=\sum_{j=0}^{r}{\bf x}_{j}^+ \Lambda_{j}|{\bar \Omega}\rangle=0.
\label{finite}\end{eqnarray}
Even though (\ref{xpm}) are valid only on the ``ground states," eqs.\
(\ref{finite}) are valid on the entire degenerate eigenspace. This can
be seen easily by applying ${\bf x}_m^-$ on the first and
${\bf x}_{m}^+$ on the second, and since all lowering (raising)
operators commute, we find these equations are valid for the entire
space generated by them. Now we can use ideas presented in Davies'
paper~\cite{Davies}. Consider the Drinfeld polynomial in (\ref{Poly}),
\begin{equation}
P(z)=\sum_{n=0}^r\Lambda_n z^{n}=\prod_{j=1}^r(z-z_j),
\quad z=t^N.
\label{drinfeld}\end{equation}
We may define, on the set of states where eqs.\ (\ref{finite}) hold,
\begin{equation}
{\bf x}_j^- =\sum_{m=1}^r z_m^{j}{\bf E}_m^-,\quad
{\bf x}_j^+=\sum_{m=1}^r z_m^{j}{\bf E}_m^+, \quad
{\bf h}_j=\sum_{m=1}^r z_m^{j}{\bf H}_m,
\label{EEH}\end{equation}
where $z_j$ are the roots of the Drinfeld polynomial. We use
(\ref{loop}), in which the operator on the left depends only on the
sum of the indices of the operators inside the commutator, to show
\begin{equation}
[{\bf E}_m^+,{\bf E}_n^-]=\delta_{m,n}{\bf H}_m,\;\;
[{\bf H}_m,{\bf E}_n^-]=2\delta_{m,n}{\bf E}_m^-,\;\;
[{\bf H}_m,{\bf E}_n^+]=-2\delta_{m,n}{\bf E}_m^+.
\label{comEH}\end{equation}
Thus, the loop algebra is decomposed into the direct sum of $r$
copies of ${\frak{sl}}_2$ algebras. Moreover, it is possible
though nontrivial to show that $({\bf E}_j^-)^2|\Omega\rangle=0$.
The degeneracy of the
eigenspace of $\tau_2(t_q)$, corresponding to the eigenvalue in
(\ref{eigen2}) for $Q=0$, which is generated by these $r$ sets of
operators of ${\frak{sl}}_2$, is indeed $2^r$.

We have chosen our loop algebra generators different from those used
in~\cite{JimboNK,NiDe,Degu2}, as we did not use the $\mathcal{L}$ matrices
used in~\cite{BPA,BS,Tara,Roan}, one of them being shown as a
``square" in Fig.~1. Rather we used the dual approach using operators
$\mathcal{U}(a,b,c,d)$, one of which is indicated in Fig.~1 by a ``star."
The operators used here in (\ref{ef}) and (\ref{u}) are more
easily seen to be lowering and raising operators.

Even though, there is ample evidence that relations (\ref{serre})
hold. Yet, the proof is still lacking. Obviously, operators
used by us are closely related to the ones used
by~\cite{Degu,JimboNK,NiDe,Degu2}. Perhaps, by mapping one to the other, a
proof of these identities may be found. This also will provide many
interesting identities of $q$-series at roots of unity.

There remains a great deal to be done for cases when $Q\ne0$.
We can show that
\begin{eqnarray}
&&[{\bf A}(t)+\omega^m{\bf D}(t)]{\bf B}_L^{(N-m)}{\bf B}_1^{(m)}=
{\bf B}_L^{(N-m)}{\bf B}_1^{(m)}[\omega^m{\bf A}(t)+{\bf D}(t)]
\nonumber\\
&&\qquad+(1\!-\!\omega)[(\omega t)^{\!-\!1}{\bf B}(t){\bf
B}_L^{(N\!-\!m)} {\bf B}_1^{(m-1)}({\bf D}_0\!-\!1)\nonumber\\
&&\qquad+\omega^m{\bf B}(t)
{\bf B}_L^{(N\!-\!m\!-\!1)}{\bf B}_1^{(m)}({\bf A}_L\!-\!1)],\nonumber\\
&&[{\bf A}(t)+\omega^m{\bf D}(t)]{\bf C}_0^{(N-m)}{\bf C}_{L-1}^{(m)}=
{\bf C}_0^{(N-m)}{\bf C}_{L-1}^{(m)}[\omega^m{\bf A}(t)+{\bf D}(t)]
\nonumber\\
&&\qquad+(1\!-\!\omega)[t\omega{\bf C}(t){\bf C}_0^{(N\!-\!m)}
{\bf C}_{L\!-\!1}^{(m-1)}({\bf D}_0\!-\!1)\nonumber\\
&&\qquad+\omega^m{\bf C}(t){\bf C}_0^{(N\!-\!m\!-\!1)}
{\bf C}_{L-1}^{(m)}({\bf A}_L\!-\!1)] .
\label{adq}\end{eqnarray}
These are related to operators that commute with
$\tau_2(t_q)\big|_Q$ given in (\ref{tauq}). Using (\ref{eigenAD}),
we find some of the eigenvectors of $\tau_2(t_q)\big|_Q$ for $Q\ne0$,
\begin{eqnarray}
&&[{\bf A}(t)\!+\!\omega^Q{\bf D}(t)]y_{Q}^-|\Omega\rangle=
\omega^Q\epsilon_{\!-Q}\, y_{Q}^-|\Omega\rangle,\quad
y_{Q}^-={\bf B}_L^{(N\!-Q)}{\bf B}_1^{(Q)},\nonumber\\
&&[{\bf A}(t)\!+\!\omega^Q{\bf D}(t)]z_{Q}^-|\bar\Omega\rangle=
\epsilon_{Q}\,z_{Q}^-|\bar\Omega\rangle,\quad
z_{Q}^-={\bf C}_0^{(N\!-Q)}{\bf C}_{L-1}^{(Q)},
\label{eigenadq}\end{eqnarray}
where $\epsilon_{Q}=
[(1\!-\!\omega t)^L\!+\!\omega^{Q}(1\!-\!t)^L]$.
This shows that $\epsilon_{Q}$ and $\omega^Q\epsilon_{-Q}$ are
eigenvalues of ${\bf A}(t)\!+\!\omega^Q{\bf D}(t)$. The eigenspaces
for $Q\ne0$ are clearly seen to be much different from the ones for
$Q=0$. For these cases, we have not yet made much progress in finding
the degeneracy of their eigenspaces, nor all the eigenvectors.

%%%%%%%%%%%%%%%%%%%%%%%%%%%%%%%%%%%%%%%%%%%%%%%%%%%%%%%%%%%%%%%%%%%%%%%%%%%
\ack The authors thank Dr.\ T.\ Deguchi for many private communications
and Dr.\ B.~M.\ McCoy for valuable criticism.

%%%%%%%%%%%%%%%%%%%%%%%%%%%%%%%%%%%%%%%%%%%%%%%%%%%%%%%%%%%%%%%%%%%%%%%%%%%

\end{document}